\documentclass{WileyMSP-template}
\usepackage{amsmath}

\newcommand{\rvline}{\hspace*{-\arraycolsep}\vline\hspace*{-\arraycolsep}}

\begin{document}

\pagestyle{fancy}

\title{Modulational Instability and Dynamical Growth Blockade in the Nonlinear Hatano-Nelson Model}

\maketitle


\author{Stefano Longhi}



\begin{affiliations}
Prof. Stefano Longhi\\
Dipartimento di Fisica, Politecnico di Milano,Piazza L. da Vinci 32, I-20133 Milano, Italy\\
\& IFISC (UIB-CSIC), Instituto de Fisica Interdisciplinar y Sistemas Complejos, E-07122 Palma de Mallorca, Spain \\
Email Address: stefano.longhi@polimi.it

\end{affiliations}



\begin{abstract}

The Hatano-Nelson model is a cornerstone of non-Hermitian physics, describing asymmetric hopping dynamics on a one-dimensional lattice, which gives rise to fascinating phenomena such as directional transport, non-Hermitian topology, and the non-Hermitian skin effect. It has been widely studied in both classical and quantum systems, with applications in condensed matter physics, photonics, and cold atomic gases. Recently, nonlinear extensions of the Hatano-Nelson model have opened a new avenue for exploring the interplay between nonlinearity and non-Hermitian effects. Particularly, in lattices with open boundary conditions, nonlinear skin modes and solitons, localized at the edge or within the bulk of the lattice, have been predicted. In this work, we examine the nonlinear extension of the Hatano-Nelson model with periodic boundary conditions and reveal a novel dynamical phenomenon arising from the modulational instability of nonlinear plane waves: growth blockade. This phenomenon is characterized by the abrupt halt of norm growth, as observed in the linear Hatano-Nelson model, and can be interpreted as a stopping of convective motion arising from self-induced disorder in the lattice.

\end{abstract}


\section{Introduction}

The Hatano-Nelson (HN) model is a non-Hermitian quantum mechanical model introduced by Naomichi Hatano and David Nelson in the 1990s \cite{r1,r2,r3}. Originally studied in the context of Anderson localization in random media under the influence of an imaginary gauge field \cite{r1,r2,r3,r4,r5,r6,r7,r8,r9,r10,r11,r12,r13,r13b,r14}, the HN model has become fundamental to understanding non-Hermitian topology \cite{r15,r16,r17,r18,r19,r20,r21,r22,r23,r24,r25,r26,r27,r27b}--a rapidly expanding field in non-Hermitian physics that investigates topological phases of matter in systems where energy is not conserved \cite{r28}. The HN model exhibits a nontrivial point-gap topology and a strong sensitivity of the energy spectrum on the boundary conditions, resulting in the non-Hermitian skin effect \cite{r17,r20,r25,r26,r27,r28,r29,r30}-- a unique phenomenon where the bulk states of a system become localized at the boundaries, defying conventional bulk-boundary correspondence seen in Hermitian systems. Introducing many-body effects and nonlinearity into the HN model leads to a variety of novel effects, which are attracting a great attention recently \cite{r31,r32,r33,r34,r35,r36,r37,r38,r39,r40,r41,r42,r43,r44,r45,r46,r47,r48,r49,r50,r51,r52,r53,r53b}. 
In particular, nonlinear extension of the original HN model incorporating a Kerr-type (focusing or defocusing) nonlinearity, has been introduced in recent works \cite{
r49,r50,r51,r52,r53,r53b}, leading to the predictions and very recent observation of nonlinear skin effect and skin solitons in systems with open boundary conditions. 
The Hermitian limit of the nonlinear HN model is the discrete nonlinear Schr\"odinger Equation (DNLSE), which is a fundamental model in nonlinear dynamics and mathematical physics describing the evolution of wave-like excitations on discrete lattices with nonlinearity in a variety of physical systems ranging from optics to Bose-Einstein condensates and lattice systems (see e.g. \cite{r54,r55,r56,r57,r58,r59,r60} and references therein). The DNLSE can be regarded as a mean field model of  the many-body Bose-Hubbard model \cite{r61}. A fundamental phenomenon in nonlinear wave dynamics is the modulational instability (MI) \cite{r55,r56,r60,r62}. It refers to the exponential growth of small perturbations in a plane wave (spatially-homogeneous) solution, leading rather generally to the formation of spatial localized structures such as solitons, breathers or localized patterns, or complex spatio-temporal dynamics.\\
In this work we investigate the phenomenon of modulational instability in the nonlinear Hatano-Nelson model under periodic boundary conditions and unveil a novel dynamical phenomenon arising from the modulational instability of nonlinear plane waves: growth blockade. This phenomenon is characterized by the abrupt halt of intensity growth, as observed in the linear Hatano-Nelson model, and can be explained as a stopping of convective motion arising from self-induced disorder in the lattice. Our results indicate that combination of nonlinearity and non-Hermitian characteristics opens up new avenues for studying wave propagation, localization, and instabilities in both classical and quantum non-Hermitian systems.

\section{Modulational instability in the nonlinear Hatano-Nelson model}

Before considering the nonlinear HN model, let us first briefly review the concept of modulational instability in the DNLSE \cite{r62}, to which the nonlinear HN reduces in the Hermitian limit. 
The DNLSE on a one-dimensional lattice reads \cite{r55,r56,r57}
\begin{equation}
i \frac{d \psi_n}{dt}= \kappa  \psi_{n+1}+\kappa  \psi_{n-1}+ \chi | \psi|^2 \psi_n
\end{equation}
for the excitation amplitude $\psi_n$ at the $n$-th lattice site, where $\kappa$ is the hopping amplitude and 
$\chi$ is the strength of the cubic (Kerr) nonlinearity. We consider a finite lattice comprising $N$ sites with periodic boundary conditions $\psi_{n+N}=\psi_n$. Under such periodic boundaries, the DNLSE admits of a family of nonlinear plane waves, which are the nonlinear extensions of the usual Bloch eigenmodes in the linear limit $\chi=0$. Such plane waves are given by
 \begin{equation}
 \psi_n(t)=A \exp[iq n-i \omega(q) t-i \theta(t) ]  \equiv \phi_n(t)
 \end{equation}
where $q$ is the wave number, which is quantized as $q=q_l= 2 l (\pi/N)$ ($l=0,1,2,..,N-1$), 
\begin{equation}
\omega(q)=2 \kappa \cos(q)
\end{equation}
is the dispersion relation of the linear lattice, and 
\[
\theta(t)=\chi |A|^2 t 
\]
is the additional nonlinear-induced phase shift.  The stability of the plane wave solution against small perturbations can be performed by a standard linear stability analysis \cite{r57,r62}, looking for a solution to Eq.(1) of the form
\begin{equation}
\psi_n(t)=[A+u_n(t)] \exp[iqn-i \omega(q)t-i \theta(t)]
\end{equation}
where $u_n(t)$ is a small perturbation. Assuming $|u_n(t)| \ll |A|$, substitution of Eq.(4) into Eq.(1) yields the following linearized equation for the perturbation amplitudes $u_n(t)$ 
\begin{equation}
i \frac{du_n}{dt}= \kappa  \exp(iq) u _{n+1}+\kappa \exp(-iq)  u_{n-1}+ (-\omega(q)+ \chi |A|^2) u_n+ \chi A^2 u_n^*.
\end{equation}
The most general solution to Eq.(5) is an arbitrary superposition of waves
\begin{equation}
u_n(t)=X \exp[iQn-i \lambda(Q)t]+ Y^* \exp[-iQn+i \lambda^*(Q)t]
\end{equation}
where $Q$ is the spatial wave number of the perturbation and $(X,Y)$, $\lambda(Q)$ are the eigenvectors and corresponding eigenvalues (Lypaunov exponents) of the $2 \times 2$ linearized matrix
\begin{equation}
M(Q)=
\left(
\begin{array}{cc}
2 \kappa \cos (q+Q)- \omega(q)+ \chi |A|^2 & \chi A^2 \\
- \chi A^{*2} & -2 \kappa \cos(q-Q)+\omega(q)- \chi |A|^2
\end{array}
\right).
\end{equation}
This yields
\begin{equation}
\lambda(Q)=-2 \kappa \sin q \sin Q \pm \sqrt{\left(4 \kappa \cos q \sin^2(Q/2)- \chi |A|^2 \right)^2- \chi^2|A|^4}.
\end{equation}
An instability arises whenever ${\rm Im}(\lambda(Q))>0$, with a MI gain given by $g(Q)={\rm Im}(\lambda(Q))$. Clearly, for a focusing nonlinearity $\chi>0$ the plane waves with wave number $q$ such that $\cos q>0$ undergo MI, while those with wave number $q$ such that $\cos q <0$ are linearly stable. The reverse occurs in the defocusing regime $\chi<0$. 
The main effects of MI is to destabilize the homogeneous intensity pattern of the plane wave leading rather generally to irregular (chaotic-like) dynamics \cite{r62} or the creation of more regular localized solutions such as train of solitons and breathing modes \cite{r57}. A few examples of the onset of MI in the DNLSE, leading to irregular spatio-temporal dynamics, are shown in Fig.1. The results are obtained by numerical integration of Eq.(1) using an accurate variable-step fourth-order Runge-Kutta method for a few different initial conditions $\psi_n(0)$ in  a lattice comprising $N=12$ sites with periodic boundary conditions. A focusing ($ \chi=1$) or defocusing ($\chi=-1$) nonlinearity has been considered. As an initial condition, we assumed a nonlinear plane wave with wave number $q$ and amplitude $A$, slightly perturbed by a small random noise, namely
\[
\psi_n(0)=A(1+r_n) \exp(iqn)
\]
where $r_n$ is a complex random variable whose real and imaginary parts are uniformly distributed in the interval $(-\delta,\delta)$ (typically we set $\delta=10^{-3}$). The results clearly show the different MI regions depending on the sign of nonlinearity $\chi$, and the existence of stable plane wave solutions.\\ 
 \begin{figure}
  \includegraphics[width=12cm]{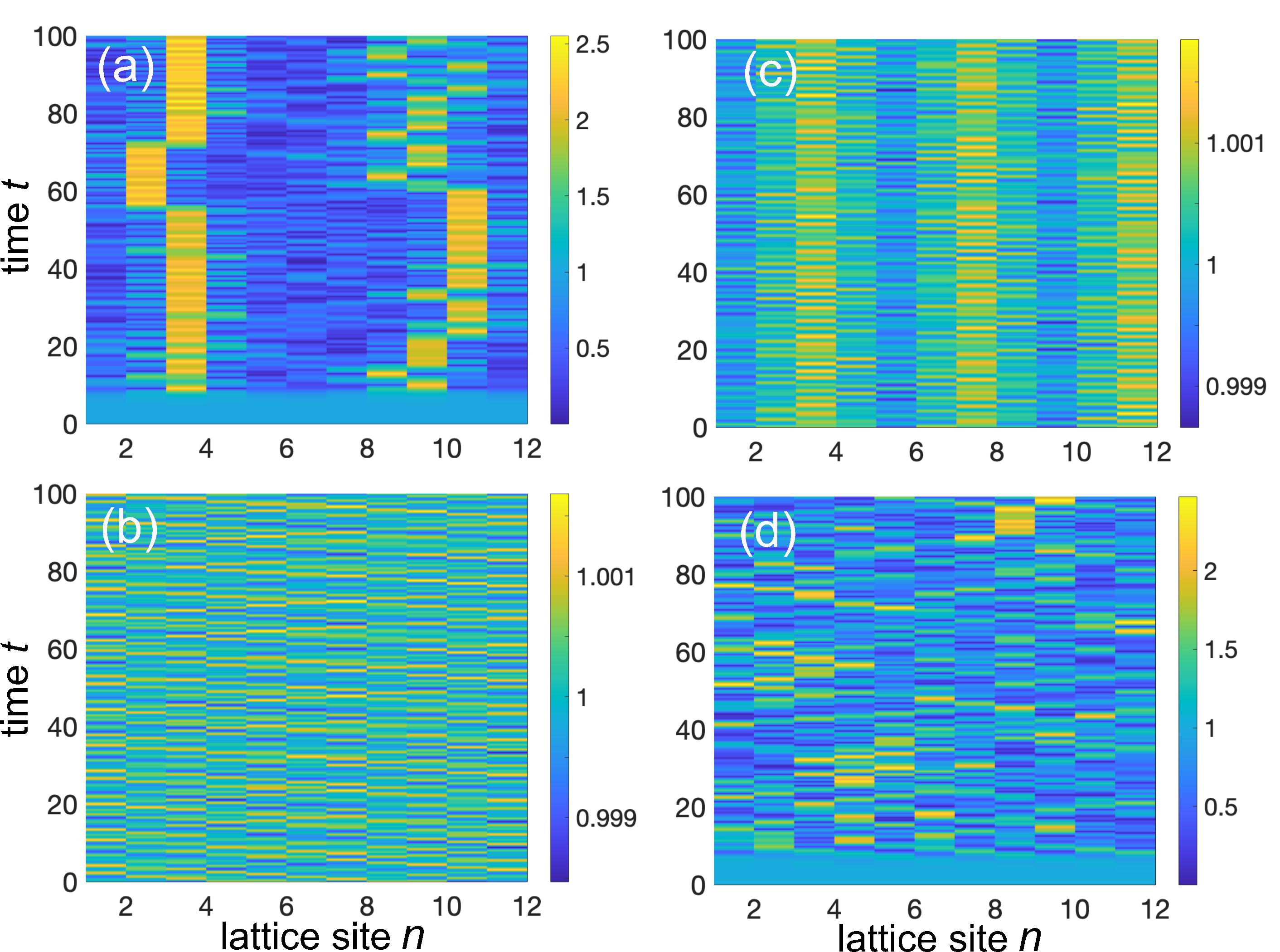}
  \caption{Onset of modulational instability in the DNLSE. (a) Numerically-computed temporal evolution of the amplitudes $|\psi_n(t)|$ on a pseudocolor map in a lattice comprising $N=12$ sites with periodic boundary conditions for hopping rate $\kappa=1$ and nonlinear parameter $\chi=1$.  The initial condition is the plane wave $\psi_n(t)=A\exp(iqn)$ with amplitude $A=1$ and spatial wave number $q=0$, perturbed by adding a small random noise. Note that the plane wave is modulationally unstable. (b) Same as (a), but for $\chi=-1$. In this case the modulational instability is absent and the plane wave is stable. (c) Same as (a) but for $q=2 \pi/3$. The plane wave is stable. (d) Same as (c) but for $\chi=-1$. The plane wave is unstable.}
\end{figure}

Let us now consider the nonlinear extension of the Hatano-Nelson model incorporating Kerr-type nonlinearity, as introduced in recent studies \cite{r49,r50, r52,r53}, which is basically obtained from the DNLSE (1) by making the left/right hopping amplitudes asymmetric via an imaginary gauge field $h$ \cite{r1}. The nonlinear HN model reads 
\begin{equation}
i \frac{d \psi_n}{dt}= \kappa \exp(h) \psi_{n+1}+\kappa \exp(-h) \psi_{n-1}+ \chi | \psi|^2 \psi_n
\end{equation}
where $h \geq 0$ is the imaginary gauge field. Clearly, in the Hermitian limit $h=0$ the above equation reduces to the DNLSE (1), whereas in the linear limit $\chi=0$ it reproduces the disorder-free HN model \cite{r1,r2,r15}. The linear HN model displays a nontrivial point-gap topology, characterized by a topological winding number \cite{r15}, and the non-Hermitian skin effect (NHSE) under open or semi-infinite boundary conditions, i.e. the localization of bulk eigenstates at the lattice edge \cite{r17,r26}. The central result in the band theory of NH systems is that the energy spectra $\sigma (H_{PBC})$, $\sigma (H_{OBC})$ and $\sigma (H_{SIBC})$ of the system Hamiltonian $H$, corresponding to periodic boundary conditions (PBC), open boundary condition (OBC), and semi-infinite boundary conditions (SIBC),  are distinct, which implies the emergence of the NHSE, topological NH edge states and the need for a non-Bloch band theory \cite{r26,r27,r27b}. Specifically, for the linear HN model: (i) $\sigma(H_{PBC})$ is a closed loop (an ellipse) in complex energy plane described by the Bloch Hamiltonian 
\[
H(q)=\kappa \exp(i q+h)+ \kappa \exp(-iq-h) 
\]
($-\pi \leq q < \pi$ is the Bloch wave number); (ii) $\sigma(H_{OBC})$ is the segment $(-2\kappa,2\kappa)$ on the real energy axis, which is always topological trivial in terms of a point gap; (iii) $\sigma(H_{SIBC})=\sigma(H_{PBC}) \bigcup B $, where $B$ is the interior of the PBC energy spectrum loop such that for $E\in B $ the winding number $W(E)$, defined by \cite{r26}
\begin{equation}
W(E)= \frac{1}{2 \pi i} \int_{-\pi}^{\pi} dq  \frac{d}{dq} \log  \det \left\{ H(q)-E \right\} \nonumber
\end{equation} 
is non vanishing. If $W(E)<0$, then $E$ is an eigenvalue of $H_{SIBC}$ of multiplicity $|W(E)|$, and the corresponding (right) eigenvectors are exponentially localized at the left edge. 
Such a result provides a bulk-boundary correspondence for the linear HN model, relating the appearance of skin edge states in a semi-infinite lattice to the topology of the PBC energy spectrum \cite{r15,r26}.
In the nonlinear extension of the HN model, the system dynamics are no longer governed by linear superposition principles but instead exhibit nonlinear effects like feedback and self-interactions. In particular, it is not possible anymore to classify edge (skin) states in terms of a topological winding number since under open boundary conditions the inclusion of Kerr-type nonlinearity alters the NHSE-induced localization leading to new boundary and bulk phenomena investigated in some recent works \cite{r49,r50, r52,r53}. While in the linear model the NHSE results in the localization of states at one edge as established by the topological winding number $W(E)$ of the PBC energy spectrum, nonlinearity can modify or redistribute the eigenstate localization \cite{r52}. Depending on the strength and nature of the nonlinearity, it can either enhance or suppress the NHSE due to the interplay between nonlinear self-trapping and linear convective drift, potentially leading to the emergence of new localized states in the bulk, such as skin solitons \cite{r53}, which clearly demonstrate a breakdown of the bulk-edge correspondence observed in the linear HN model. Interestingly, under open boundary conditions the nonlinear HN model can be mapped onto the Hermitian DNLSE with spatially-inhomogeneous nonlinearity via the non-unitary gauge transformation $ \psi_n(t)  \rightarrow \psi_n(t) \exp(-hn)$ \cite{r53}.\\
Unlike recent studies \cite{r49,r50,r51,r52,r53}, here we focus our attention to a fully different regime, corresponding to a lattice with periodic boundary conditions where such a mapping is prevented. 
As we will show below, in this case the nonlinearity drives the system into an Anderson-type disordered phase triggered by the modulational instability of nonlinear plane waves. In fact,  under periodic boundary conditions, i.e. $\psi_{n+N}(t)=\psi_n(t)$,
Eq.(9) admits of the following family of {\em exact nonlinear} plane wave solutions 
 \begin{equation}
 \psi_n(t)=A \exp[iq n-i \omega(q) t-i\theta(t)] \equiv \phi_n(t)
 \end{equation}
where $q$ is the wave number, which is quantized as $q=q_l= 2 l (\pi/N)$ ($l=0,1,2,..,N-1$), 
\begin{equation}
\omega(q)=2 \kappa \cosh(h+iq)
\end{equation}
is the dispersion relation of plane waves of the linear HN model, and
\begin{equation}
\theta(t)= \chi |A|^2 \int_0^{t} d \tau \exp[ 2 {\rm Im}(\omega(q)) \tau]= \frac{\chi |A|^2}{2 {\rm Im}(\omega(q))} \left[ \exp(2 {\rm Im}(\omega(q)) t) -1\right]
\end{equation}
is the nonlinear time-varying phase shift accumulated during the evolution. Clearly, in the Hermitian (conservative) limit $h=0$ such plane waves reduce to Eq.(2) of the DNLSE. For $h>0$, the frequency $\omega(q)$ is complex, corresponding to wave damping for $q<0$ and wave amplification for $q>0$, with the largest growth rate attained at $q = \pi/2$. Unlike the Hermitian limit of the DNSE, for $h>0$ the plane solutions are exponentially growing or damped along the propagation, with an exponential growth of the phase term $\theta(t)$ according to Eq.(12). Only for $q=0, \pi$ the growth rate ${\rm Im}(\omega)$ vanishes and $\theta(t)$ varies linearly with time $t$ like in the Hermitian limit. Hence, the usual concept of "linear stability" and "modulational instability" of nonlinear plane wave solutions for the DNSLE, established by the behavior of Lyapunov exponents (eigenvalues) of linearized equations [Eq.(8)], should be properly extended when $h$ is nonvanishing.  To test the stability of the plane wave solution (10), let us consider the perturbed solution
\begin{equation}
\psi_n(t)= \phi_n(t)\left[ 1+u_n(t) \right]
\end{equation}
where $u_n(t)$ is a small perturbation, i.e. $|u_n(t)| \ll 1$. Substitution of Ansatz (13) into Eq.(9) and after linearization yields the following evolution equation for the perturbations $u_n(t)$
\begin{equation}
i \frac{du_n}{dt}  =  -\omega(q) u_n+ \kappa \exp(h+iq) u_{n+1}+\kappa \exp(-h-iq) u_{n-1}+ \dot{\theta} (u_n+u_n^*)
\end{equation}
where $\dot{\theta}=(d \theta/ dt)= \chi |A|^2 \exp[2 {\rm Im }(\omega(q)) t]$.
The plane wave solution (10) is said to be linearly stable whenever, for any initial perturbation $u_n(0)$ at time $t=0$ with $|u_n(0)| \ll 1$, as $t \rightarrow \infty$ 
$|u_n(t)|$ remains much smaller than one. The solve the linearized equation (14), let us introduce the Ansatz
\begin{equation}
u_n(t)=F(t) \exp (iQn)+G^*(t) \exp(-iQn),
\end{equation} 
where $Q$ is the spatial wave number of the perturbation. Substitution Eq.(15) into Eq.(14) yields the following coupled equations for the complex amplitudes $F(t)$ and $G(t)$
\begin{eqnarray}
i \frac{dF}{dt} &= &[ \omega(q+Q) -\omega(q) ]F + \dot{\theta} ( F+  G )\\
- i \frac{dG}{dt} &= & [\omega^*(q-Q) -\omega^*(q)] G + \dot{\theta}(G+F)
\end{eqnarray}
To establish the asymptotic behavior of the amplitudes $F$ and $G$ as $t \rightarrow \infty$,
we should distinguish three cases, depending on the value of the wave number $q$ of the nonlinear wave mode.\\
(i) For a wave number $-\pi < q<0$, corresponding to ${\rm Im} (\omega(q))<0$, the nonlinear wave is damped,  i.e. $\psi_n(t) \rightarrow 0$ as $t \rightarrow \infty$, 
and in the large $t$ limit the evolution equations for the amplitudes $F(t)$ and $G(t)$ [Eqs.(16) and (17)] are uncoupled and their growth rates are given by the imaginary parts of $\omega(q \pm Q)$. Clearly, there are intervals of wave number $Q$ of spatial perturbations such that ${\rm Im}(\omega(q \pm Q))> {\rm Im}(\omega(q))$. This means that, as $t \rightarrow \infty$, the perturbation amplitude $|u_n(t)|$ exponentially grows from a small initial value, indicating that the nonlinear plane wave with wave number $q$ is unstable.\\
(ii) For $q=0, \pi$, the frequency $\omega(q=0)=\pm 2 \kappa \cosh (h)$  of the nonlinear plane wave is real, $\dot{\theta}$ is constant in time and a standard linear stability analysis based on eigenvalue analysis, as in the DNLSE case, can be performed in Eqs.(16) and (17). The analysis indicates that the nonlinear wave is modulationally unstable is some intervals of the perturbation wave numbers $Q$.\\
(iii) For  a wave number $0<q<\pi$, corresponding to ${\rm Im} (\omega(q))>0$, the nonlinear wave is amplified and it turns out that at any wave number $Q$ of spatial perturbations the amplitudes $G(t)$, $F(t)$ are unbounded. This result is nontrivial and involves a detailed asymptotic analysis of the non-autonomous system Eqs.(16) and (17), which is presented in Appendix A.\\
The above results indicate that, unlike the NLDSE, in the nonlinear HN model any nonlinear plane wave undergoes MI, regardless the value of the wave number $q$. Remarkably, such a result holds even for plane waves displaying an exponential growth, i.e. with ${\rm Im}(\omega(q))>0$, which are unstable to any spatial wave number perturbation.

\section{Dynamical growth blockade}

The linear stability analysis of plane wave solutions in the nonlinear HN model reveals that these modes are always modulationally unstable, unlike the conservative limit of the DNSLE ($h=0$). However, this analysis does not provide insight into the system's behavior as the instability evolves. To address this, numerical simulations of Eq.(9) are required. Assuming as an initial condition a plane wave with spatial wave number $q$, slightly perturbed by a small random fluctuations of the amplitudes, an instability  clearly sets in leading to complex spatio-temporal patterns. Examples of numerical results are shown in Figs.2,3 and 4 for both positive and negative sign of the nonlinear coefficient $\chi$. The figures illustrate the typical evolution of the wave amplitude $|\psi_n(t)|$ on a pseudocolor map, and corresponding behavior of total intensity $P(t)$, defined by
\begin{equation}
P(t)=\sum_{n=1}^N | \psi_n(t)|^2,
\end{equation}
 \begin{figure}
  \includegraphics[width=12cm]{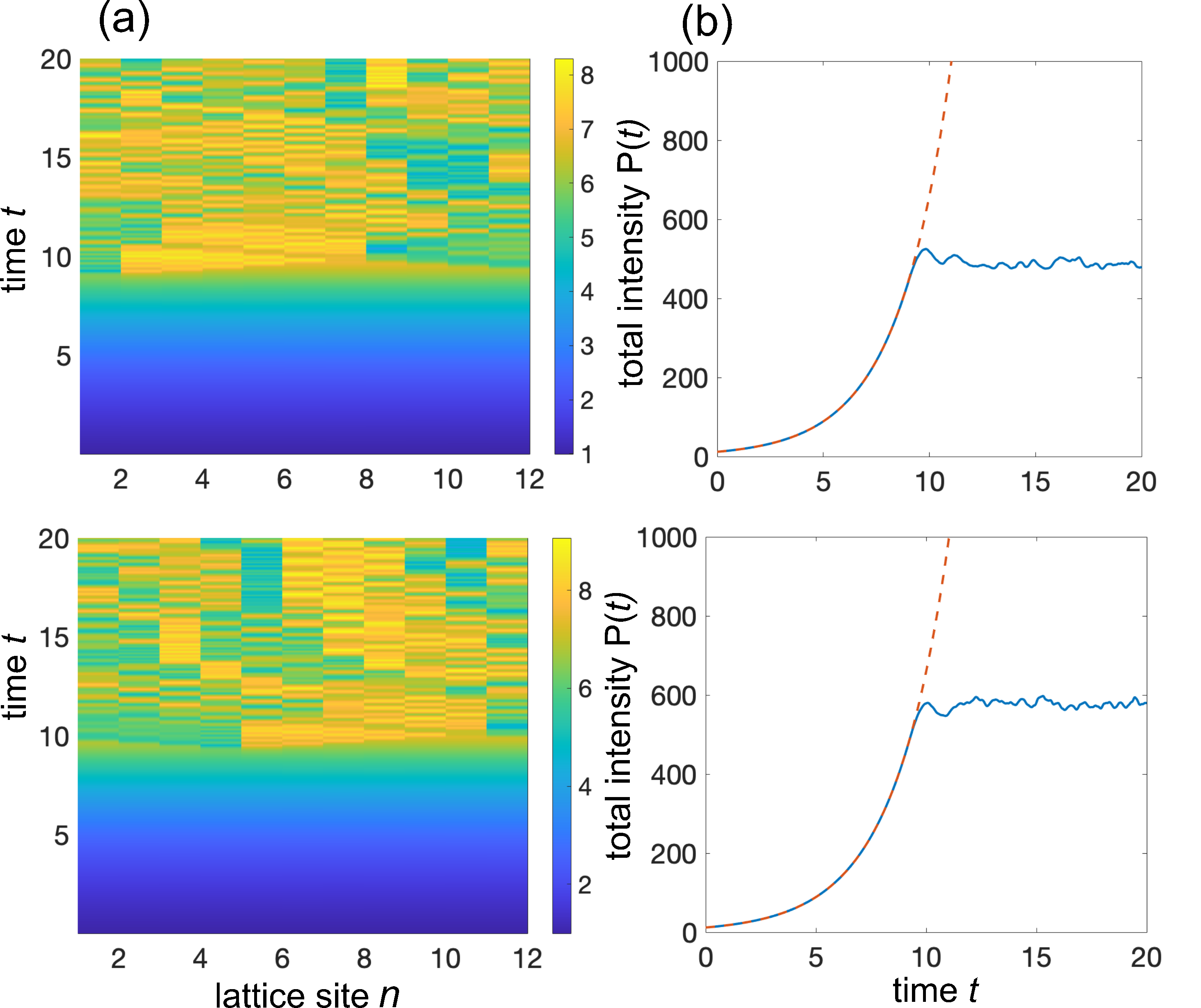}
  \caption{Illustration of the growth blockade effect. (a) Numerically-computed temporal evolution of the amplitudes $|\psi_n(t)|$ on a pseudocolor map in a lattice comprising $N=12$ sites with periodic boundary conditions for parameter values $\kappa=1$ (hopping rate), $h=0.1$ (imaginary gauge field) and for a nonlienar coefficient $\chi=1$ (upper panels) and $\chi=-1$ (lower panels).  The initial condition is the plane wave $\psi_n(t)=A\exp(iq_0n)$ with amplitude $A=1$ and spatial wave number $q_0=\pi/2$, perturbed by adding a small random noise. (b) Corresponding temporal evolution of the total excitation intensity $P(t)=\sum_n | \psi_n(t)|^2 $ (solid curve). The dashed curve is the behavior of $P(t)$ of the nonlinear plane wave solution, given by the exponential law $P(t)=N|A|^2 \exp \left\{ 2 {\rm Im}(\omega (q_0) ) t \right\}$. Note that the exponential growth of the total excitation intensity is arrested after an initial transient, and $P(t)$ settles down to a nearly constant value with small fluctuations for times $t  > \sim 10$. The arrest of the secular intensity growth is associated to the emergence of an irregular spatial distribution of $\psi_n(t)$.}
\end{figure}

 \begin{figure}
  \includegraphics[width=12cm]{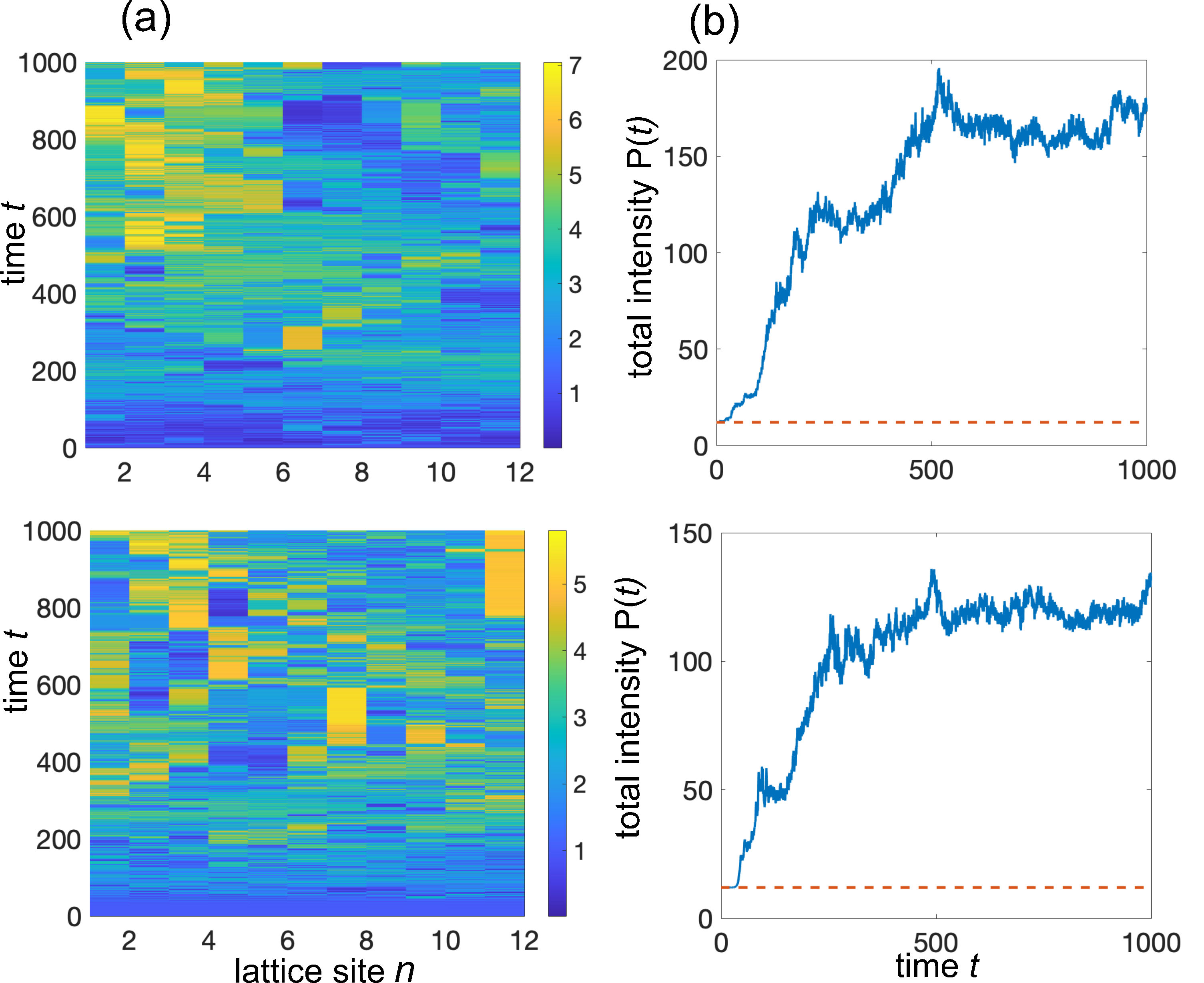}
  \caption{Same as Fig.2, but for $q_0=0$.}
\end{figure}

 \begin{figure}
  \includegraphics[width=12cm]{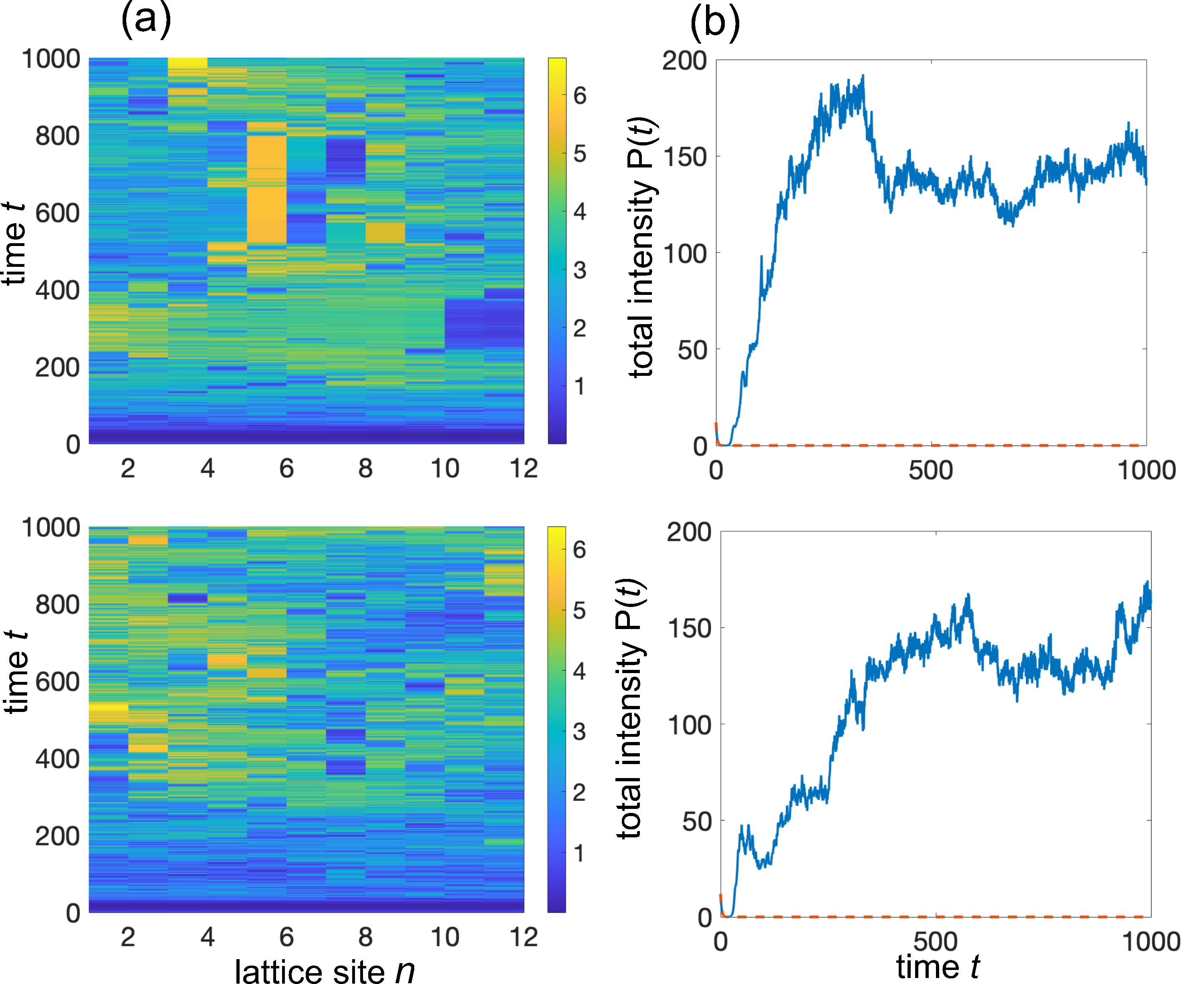}
  \caption{Same as Fig.2, but for $q_0=-\pi$.}
\end{figure}

as obtained by numerical integration of Eq.(9) using an accurate variable-step fourth-order Runge-Kutta method for a few different initial conditions $\psi_n(0)$. A lattice comprising $N=12$ sites with periodic boundary conditions has been assumed in the numerical simulations, and both the focusing ($ \chi=1$) and defocusing ($\chi=-1$) nonlinearities have been considered. As an initial condition, we assumed a nonlinear plane wave with wave number $q$ and amplitude $A$, slightly perturbed by a small random noise, namely
\begin{equation}
\psi_n(0)=A(1+r_n) \exp(iqn)
\end{equation}
where $r_n$ is a complex random variable whose real and imaginary parts are uniformly distributed in the interval $(-\delta,\delta)$ (typically we set $\delta=10^{-3}$).
A general characteristic of the clean HN model under periodic boundary conditions --corresponding to the linear limit,  $ \chi=0$, in Eq. (9) -- is that, for a wide range of initial conditions (including random ones), the total excitation intensity in the system $P(t)$
displays a secular growth in time, with a largest growth rate given by $ 2 \kappa \sinh h$, which is the largest value of ${\rm Im}(\omega(q))$ attained at $q = \pi/2$. One might naively expect a similar behavior in the nonlinear regime ($ \chi \neq 0$), since the nonlinear term is conservative and does not directly affect the power balance in the system. However, this is not the case: after an initial transient, a blockade of the secular intensity growth is observed, as shown in Figs.2,3 and 4. Interestingly, when the initial condition is the plane wave corresponding to the wave number $q=\pi/2$ with the largest growth rate $2 \kappa \sinh h$, the temporal evolution of $P(t)$ shows an initial exponential growth of the nonlinear plane wave mode, followed by an abrupt halt of growth, as shown in Fig.2(b). A natural question arises: what is the physical origin of the growth blockade in the nonlinear HN model observed in the numerical simulations?  Broadly speaking, this growth arrest can be understood as a stopping of convective motion due to self-induced disorder in the system originating from the modulation instability. In fact, in the clean (disorder-free) linear HN model the growth of $P(t)$ is the result of a {\em convective} instability in the bulk of the lattice \cite{r25,r63,r64,r65}, associated to a biased drift of the excitation: preventing the biased drift, for example by an edge boundary or by strong impurities or strong enough on-site potential disorder in the lattice, halt the secular growth of the total excitation intensity $P(t)$.  For example, when strong on-site potential disorder $V_n$ is added to the linear HN model, i.e. when we consider the model 
\begin{equation}
i \frac{d \psi_n}{dt}= \kappa  \psi_{n+1}+\kappa  \psi_{n-1}+ V_n \psi_n,
\end{equation}
owing to Anderson localization the drift along the lattice is prevented and the corresponding growth of excitation is arrested. Such an effect is well understood in the linear HN model with disorder and related to the dubbed {\em non-Hermitian Anderson localization transition} predicted in the early studies of the HN model \cite{r1,r2,r3,r4,r5,r6,r7,r8,r9,r10,r11}. In the nonlinear HN model, there is not any on-site potential disorder in the system nor edges, however a kind of on-site potential disorder is {\em self-induced} by the modulational instability of the nonlinear plane wave solution. In fact, the MI makes the local intensity term $V_n=|\psi_n|^2$ in Eq.(9) basically equivalent to a self-induced spatially-inhomogeneous and irregular on-site potential, which thus prevents drift and amplification of small-amplitude excitation along the nonlinear lattice. To check the significance of such a physical interpretation, let $\psi_n(t)= \phi_n(t)$ be a spatially-inhomogeneous (and irregular) solution to Eq.(9), as observed in Fig.2 after the initial transient growth, and let us consider the evolution of a small perturbation $u_n(t)$ added to $\phi_n(t)$. After linearization, the perturbation $u_n(t)$ satisfies the linearized system
\begin{equation}
i \frac{du_n}{dt}= \kappa \exp(h) u_{n+1}+ \kappa \exp(-h) u_{n-1} +2V_n(t) u_n+W_n(t) u_n^*
\end{equation}
where we have set
\begin{equation}
V_n(t) \equiv \chi | \phi_n(t)|^2 \; ,\;\; W_n(t) \equiv \chi \phi_n^2(t)
\end{equation}
which act as effective potentials for the the perturbation wave amplitude $u_n(t)$.
For an irregular solution $\phi_n(t)$, as observed in the numerical simulations of Fig.1 after the initial transient growth, the two effective potentials are spatially irregular and time varying. To provide qualitative insights on the growth blockade effect,  we can neglect rapid time oscillations of the effective potentials and replace them in Eqs.(21) and (22) by the their local time average
\begin{equation}
\overline{V}_n=\frac{\chi}{\Delta t} \int_t^{t+\Delta t} d \tau V_n(\tau) = \frac{\chi}{\Delta t} \int_t^{t+\Delta t} d \tau | \phi_n(\tau)|^2  \; , \;\;\  \overline{W}_n=\frac{\chi}{\Delta t} \int_t^{t+\Delta t} d \tau W_n(\tau) 
= \frac{\chi}{\Delta t} \int_t^{t+\Delta t} d \tau  \phi_n^2(\tau)  
\end{equation}
where $\Delta t$ is the time interval of averaging. For large enough $\Delta t$, the two averaged potentials are slowing varying with time $t$ and can be assumed to be locally constant, as shown as an example in Fig.5(a).  Under such approximation, the growth of perturbations $u_n(t)$ in time is established by the position in complex plane of the $2N$ eigenvalues $\lambda$ of the following $2N \times 2N$ matrix

 \begin{figure}
  \includegraphics[width=12cm]{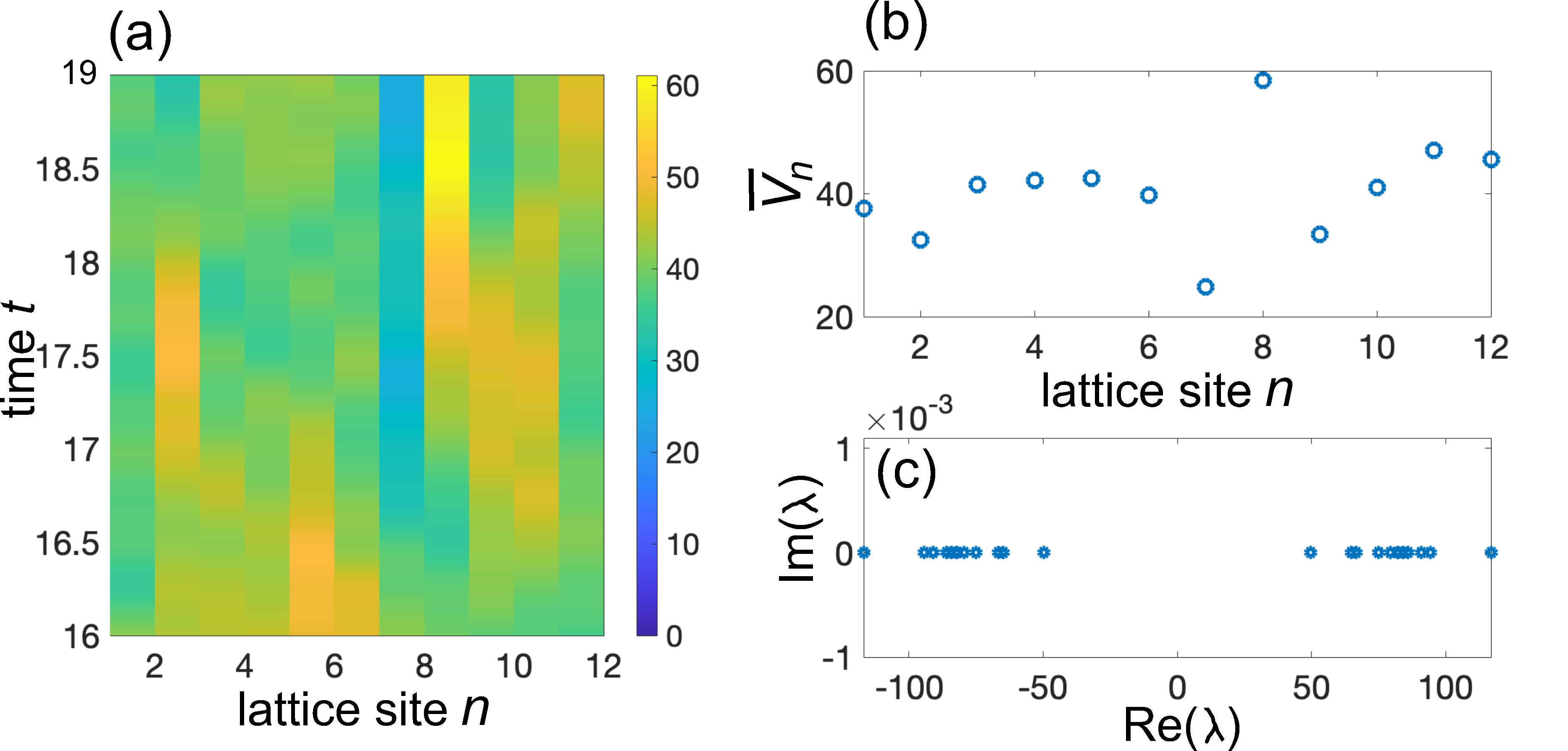}
  \caption{(a) Numerically-computed behavior of the local averaged potential $\overline{V}_n(t)=(\chi/ \Delta t) \int_t^{t+ \Delta t} |\phi_n(\tau)|^2 d \tau$ versus time $t$, corresponding to the solution $\phi_n(t)$ shown in Fig.2 with $\chi=1$, on a pseudo color map. Average time interval is $\Delta t=1$. (b) Detailed behavior of $\overline{V}_n$ for time $t=19$. (c) Eigenvalue spectrum of the matrix $\mathcal{S}$ with potential $\overline{V}_n$ plotted in panel (b). Note that the imaginary part of all eigenvalues vanishes.}
\end{figure}
\[
\mathcal{S}=\begin{pmatrix}
 \mathcal{A}
  & \rvline & \mathcal{B} \\
\hline
 - \mathcal{B}^* & \rvline &
  -\mathcal{A}
\end{pmatrix}
\]
  where the $N \times N$ matrix $\mathcal{A}$ is given by 
   \[
   \mathcal{A}=
   \begin{pmatrix}   
  \overline{V}_1 & \kappa \exp(h) & 0 & ... & 0 & 0 &  \kappa \exp(-h) \\
  \kappa \exp(-h) & \overline{V}_2 & \kappa \exp(h)& ... & 0 & 0 & 0 \\
 0 & 0  & 0 & ... & 0 & 0 \\
 0 & 0 &0 & ... & \kappa \exp(-h) & \overline{V}_{N-1} &  \kappa \exp(h) \\
 \kappa  \exp(h) & 0 &0 & ... & 0 & \kappa \exp(-h) & \overline{V}_{N}  
   \end{pmatrix} ,
   \]
   $\mathcal{B}$ is the $N \times N$ diagonal matrix defined by $B_{n,m}=\overline{W}_n \delta_{n,m}$, and $\mathcal{B}^*$ is the element-wise complex conjugate of $\mathcal{B}$. A typical eigenvalue spectrum $\lambda$ in complex plane is shown in Fig.5(c), indicating that all states are marginally stable, i.e. the imaginary part of $\lambda$ is vanishing. Hence small-amplitude perturbations cannot secularly grow once the self-induced potentials $V_n$ and $W_n$ have been created. Such an analysis is of course not rigorous, since it disregard time fluctuations of the self-induced disordered potentials, however it is capable of catching the main physics underlying the growth blockade effect as observed in numerical simulations.
  
\section{Conclusion}
In this work, we have investigated modulational instability in the nonlinear extension of the Hatano-Nelson model under periodic boundary conditions, unveiling a novel dynamical phenomenon: growth blockade. In contrast to the unbounded growth of the total excitation intensity seen in the linear HN model -- where an unsaturated convective instability dominates -- in the nonlinear regime, we observe a suppression of this expected growth after an initial transient phase. This growth arrest can be understood as a self-induced suppression of transport in the lattice, driven by dynamically induced disorder, which is a hallmark of the HN model with strong disorder or edges. This mechanism is fundamentally different from the growth clamping of modulation instability of planes waves seen in other conservative or dissipative systems, where the instability is stopped by some gain saturation mechanisms in the system leading to pattern formation or localized structures \cite{r66}.

Our results highlight the critical role played by the interplay between nonlinearity and non-reciprocal transport in shaping the system's behavior. The nonlinear HN model introduces entirely new dynamical regimes that are absent in both the linear HN model and more conventional nonlinear systems like the discrete nonlinear Schr\"odinger equation. Specifically, the combination of Kerr-type nonlinearity and non-reciprocal transport inherent in the HN model produces a range of complex behaviors, including the nonlinear skin effect and the formation of skin solitons in systems with open boundary conditions, which have been recently predicted and observed in related works. The observed growth blockade in periodic settings adds another layer to our understanding of these systems.

This work serves as a foundation for further exploration of nonlinear non-Hermitian systems, with potential applications in fields such as optics, photonics, and quantum non-Hermitian many-body physics in the mean-field regime. 


\medskip
\textbf{Conflict of Interest} \par
The author declares no conflict of interest.

\medskip
\textbf{Data Availability Statement}
The data that support the findings of this study are available from the corresponding author upon reasonable request. \par

\medskip
\textbf{Acknowledgements} \par 
The author acknowledges the Spanish State Research Agency, through
the Severo Ochoa and Maria de Maeztu Program for Centers 
and Units of Excellence in R\&D (Grant No. MDM-2017-
0711).

\medskip

%

\appendix

\section{Appendix A: Stability analysis of nonlinear plane waves}
In this Appendix it is shown that the nonlinear wave  given by Eq.(4), for a spatial wave number $0<q< \pi$ corresponding to an amplified wave (${\rm Im}(\omega(q))>0$), is modulationally unstable for any spatial wave number $Q$ of the perturbation. To this aim, let us rewrite the coupled equations (16) and (17), describing the evolution of the amplitudes $F(t)$ and $G(t)$ of the perturbation, after introduction of the stretched time variable $T$, defined via the relation
\begin{equation}
T=\int_0^t d \tau \exp[2 {\rm Im}(\omega(q)) \tau]=\frac{\exp[2 {\rm Im}(\omega(q)) t] -1}{2 {\rm Im} (\omega(q))}. \label{eq19}
\end{equation}
Taking into account that $d/dt= \exp[2 {\rm Im} (\omega(q)) t ]( d/ d T)$, Eqs.(16) and (17) take the form
\begin{eqnarray}
i \frac{dF}{d T} & = & \epsilon_1(T) F+ \sigma (F +G) \label{eq20}\\
i \frac{dG}{d T} & = & \epsilon_2(T) F- \sigma (F +G) \label{eq21}
\end{eqnarray}
where we have set $\sigma \equiv \chi |A|^2$ and 
\begin{eqnarray}
\epsilon_1(T) & = &  \frac{\omega(q+Q)-\omega(q)}{1+ 2 {\rm Im}(\omega(q)) T} \\
\epsilon_2(T) & = &  \frac{\omega^*(q)-\omega^*(q-Q)}{1+ 2 {\rm Im}(\omega(q)) T}.
\end{eqnarray}
Let us note that $\epsilon_{1,2}(T)   \sim 1/T \rightarrow 0$ as $T \rightarrow \infty$ for any value of the spatial perturbation wave number $Q$. Let us first consider the case $Q=0$, i.e. the long-wavelengh limit ($Q \rightarrow 0$) of the perturbation. In this case $\epsilon_{1,2}(T)=0$ and one obtains the autonomous linear system
 \begin{eqnarray}
i \frac{dF}{d T} & = &  \chi |A|^2 (F +G) \label{eq22}\\
i \frac{dG}{d T} & = & - \chi |A|^2 (G + F) \label{eq23}.
\end{eqnarray}
 which can be solved by standard methods. Since the matrix of such a linear system displays two vanishing eigenvalues corresponding to an exceptional point, for rather arbitrary initial conditions it turns out that the amplitudes $F(T)$ and $G(T)$ are unbounded and grow as $\sim T$. In physical time $t$, according to Eq.(\ref{eq19}) this means that the amplitudes $F$ and $G$ of the perturbation grow as $ \sim \exp[2 {\rm Im}(\omega(q)) t]$,  indicating that the nonlinear plane wave solution $\phi_n(t)$ is unstable.\\
 Let us now consider the case of a spatial perturbation with $Q \neq 0$. In this case $\epsilon_{1,2}(T)$ are distinct and complex, vanishing as $T \rightarrow \infty$, i.e. the exceptional point of the linear system is attained asymptotically in time. To study the  asymptotic behavior of the solutions in this case, we note that Eqs.(\ref{eq21}) and (\ref{eq22}) can be regarded as the equations of a two-level system described by the time-dependent non-Hermitian Hamiltonian 
 \begin{equation}
 \mathcal{H}(T)=
 \left(
 \begin{array}{cc}
 \epsilon_1(T)+ \sigma & \sigma \\
 -\sigma & \epsilon_2(T)- \sigma
 \end{array}
 \right)
 \end{equation}
 for which a non-adiabatic asymptotic analysis can be performed, similar to the one developed to study exceptional point encircling (see e.g. \cite{adia1,adia2,adia3,adia4}).  To this aim, let us indicate by $|u_{1,2}(T) \rangle$ and 
 $| u_{1,2}^{\dag}(T) \rangle$ the instantaneous right and left eigenvectors, respectively, of  $\mathcal{H}(T)$, and by $\lambda_{1,2}(T)$ the corresponding instantaneous eigenvalues,  i.e. 
 $\mathcal{H} |u_{1,2}(T) \rangle=\lambda_{1,2}(T) | u_{1,2}(T) \rangle$ and $\mathcal{H}^{\dag} | u_{1,2}^{\dag}(T) \rangle=\lambda^*_{1,2}(T) | u^{\dag}_{1,2}(T) \rangle$, with the normalization conditions $\langle u^{\dag}_{i}| u_{k} \rangle= \delta_{i,k}$ ($i,k=1,2$). It can be readily shown that, in the large $T$ limit, at leading order one has
 \begin{equation}
 \lambda_{1,2}(T)= \pm \sqrt{ \sigma} \sqrt{\epsilon_1(T)-\epsilon_2(T)} \sim \pm \frac{\Theta}{\sqrt{T}} \label{eig}
 \end{equation}
 for the instantaneous eigenvalues, where 
 \[
 \Theta \equiv \sqrt{\sigma} \sqrt{\frac{\omega(q+Q)-\omega(q)-\omega^*(q-Q)+\omega^*(q)}{ 2 {\rm Im}(\omega(q))}},
 \] 
 and
 \begin{equation}
 |u_{1,2} \rangle=
 \left(
 \begin{array}{c}
 \sigma \\
 \lambda_{1,2}- \sigma
 \end{array}
 \right) \; ,\;\;
 |u_{1,2}^{\dag}
 \rangle=
 \frac{1}{2 \sigma \lambda_{1,2}^*}
 \left(
 \begin{array}{c}
 \lambda_{1,2}^{*} + \sigma \\
 \sigma
 \end{array}
 \right) \label{vectors}
 \end{equation}
 for the corresponding right and left eigenvectors. Note that, according to Eq.(\ref{eig}), the two eigenvalues $\lambda_1$ and $\lambda_2=-\lambda_1$ are complex with opposite imaginary parts.  
 Without loss of generality, we will assume $\lambda_1$ as the dominant eigenvalue, i.e. ${\rm Im}(\lambda_1)>0$ and ${\rm Im}(\lambda_2)<0$, corresponding to ${\rm Im} (\Theta)>0$. 
 Let us  write the solution to Eqs.(\ref{eq20}) and (\ref{eq21}) in the adiabatic basis using the Ansatz
 \begin{equation}
 \left(
 \begin{array}{c}
 F(T) \\
 G(T)
 \end{array}
 \right)= \alpha_1(T) \exp \left[ -i \int^T dt \lambda_1(t) \right] |u_1(T) \rangle + \alpha_2(T) \exp \left[ -i \int^T dt \lambda_2(t) \right] | u_2(T) \rangle. \label{eq32}
 \end{equation}
Substitution of Eq.(\ref{eq32}) into Eqs.(\ref{eq20}) and (\ref{eq21}) yields the following coupled equations for the adiabatic amplitudes $\alpha_1(T)$ and $\alpha_2(T)$
\begin{eqnarray}
\frac{d \alpha_1}{dT} & = &- \alpha_1 \langle u^{\dag}_1| \frac{d u_1}{dT} \rangle -\alpha_2  \langle u^{\dag}_1 | \frac{d u_2}{dT} \rangle \exp \left[  i \int^T dt (\lambda_1-\lambda_2)  \right]  \label{adia1}\\
\frac{d \alpha_2}{dT} & = &- \alpha_2 \langle u^{\dag}_2| \frac{d u_2}{dT} \rangle -\alpha_1  \langle u^{\dag}_2 | \frac{d u_1}{dT} \rangle \exp \left[  -i \int^T dt (\lambda_1-\lambda_2)  \right] \label{adia2}
\end{eqnarray}
which are exact at this stage. Using Eq.(\ref{vectors}), Eqs.(\ref{adia1}) and (\ref{adia2}) ca be cast in the form
\begin{eqnarray}
\frac{d \alpha_1}{dT} & = &- \frac{\dot{\lambda_1}}{2 \lambda_1} \alpha_1  - \frac{\dot{\lambda_2}}{2 \lambda_1} \alpha_2   \exp \left[  i \int^T dt (\lambda_1-\lambda_2)  \right]  \label{uff1}\\
\frac{d \alpha_2}{dT} & = &-  \frac{\dot{\lambda_2}}{2 \lambda_2}\alpha_2 - \frac{\dot{\lambda_1}}{2 \lambda_2} \alpha_1  \exp \left[  -i \int^T dt (\lambda_1-\lambda_2)  \right] \label{uff2}
\end{eqnarray}
where the dot indicates the derivative with respect to $T$, i.e. $\dot{\lambda}_{1,2=} (d \lambda_{1,2} /dT)$.
Note that, using Eq.(\ref{eig}) one has asymptotically
\begin{equation}
\exp \left[  -i \int^T dt (\lambda_1-\lambda_2)  \right] \sim \exp \left( -4i \Theta \sqrt{T}  \right)
\end{equation}
which grows in time as $ \sim \exp [ 4 {\rm Im} (\Theta) \sqrt{T}]$. To determine the asymptotic behavior of amplitudes $\alpha_{1,2}(T)$ in the large $T$ limit, one can make the assumption, to be justified {\em a posteriori}, that the second term on the right hand side of Eq.(\ref{uff1}) is negligible, i.e. that $\alpha_2(T)$ can grow but slower than $\sim \exp [ 4 {\rm Im} (\Theta) \sqrt{T}]$. Under this assumption Eq.(\ref{uff1}) can be solved yielding $\alpha_1(T)=\lambda_1^{-1/2}$, which secularly grows in time as $\sim T^{1/4}$. From Eq.(\ref{uff2}), one can then calculate the asymptotic long-time behavior of $\alpha_2(T)$ by neglecting the first term on the right hand side, and the corresponding solution grows in time as $ \sim (1/T^{1/4})   \exp [ 4 {\rm Im} (\Theta) \sqrt{T}]$. Such a growth is smaller than $ \sim \exp [ 4 {\rm Im} (\Theta) \sqrt{T}]$, consistently with the initial trial assumption.  To sum up, the asymptotic analysis indicates that in the large $T$ limit the behavior of $F(T),G(T)$ is dominated by the first term on the right hand side of Eq.(\ref{eq32}), and they grow in time asymptotically as $|F|, |G| \sim T^{1/4} \exp [ 2 {\rm Im} (\Theta) \sqrt{T}]$. This proves that the nonlinear plane wave solution is modulationally unstable for $Q \neq 0$ as well.







\end{document}